\documentstyle[12pt]{article}
\newcommand{\bce}{\begin{center}}
\newcommand{\ece}{\end{center}}
\newcommand{\beq}{\begin{equation}}
\newcommand{\eeq}{\end{equation}}
\newcommand{\bea}{\vspace{0.25cm}\begin{eqnarray}}
\newcommand{\eea}{\end{eqnarray}}

\newcommand{\ba}{\begin{array}}
\newcommand{\ea}{\end{array}}

\newcommand{\doublespace}{
    \renewcommand{\baselinestretch}{1.6}\large\normalsize}

\def\lsim{\mathrel{\rlap{\lower4pt\hbox{\hskip1pt$\sim$}}
    \raise1pt\hbox{$<$}}}	  
\def\gsim{\mathrel{\rlap{\lower4pt\hbox{\hskip1pt$\sim$}}
    \raise1pt\hbox{$>$}}}	  

\def\Pom{{\bf I\!P}}

\def\lsim{\mathrel{\rlap{\lower4pt\hbox{\hskip1pt$\sim$}}
    \raise1pt\hbox{$<$}}}         
\def\gsim{\mathrel{\rlap{\lower4pt\hbox{\hskip1pt$\sim$}}
    \raise1pt\hbox{$>$}}}         

\def\Pom{{\bf I\!P}}

  \advance\oddsidemargin  by -1in
  \advance\evensidemargin by -1in
  \advance\topmargin      by -0.5in
\def\lsim{\mathrel{\rlap{\lower4pt\hbox{\hskip1pt$\sim$}}
    \raise1pt\hbox{$<$}}}         
\def\gsim{\mathrel{\rlap{\lower4pt\hbox{\hskip1pt$\sim$}}
    \raise1pt\hbox{$>$}}}         

\def\Pom{{\bf I\!P}}
\def\beq{\begin{equation}}
\def\endeq{\end{equation}}
\def\arr{\begin{eqnarray}}
\def\endarr{\end{eqnarray}}
\makeindex

\doublespace

\begin{document}


\phantom{.}{\bf \Large \hspace{10.0cm} KFA-IKP(Th)-1994-16 \\
\phantom{.}\hspace{11.9cm}10 March   1994\vspace{0.4cm}\\ }

\begin{center}
{\bf\sl \huge Splitting the pomeron into two jets: \\
a novel process at HERA}
\vspace{0.4cm}\\
{\bf \large
N.N.~Nikolaev$^{a,b}$  and B.G.~Zakharov$^{b,c}$
\bigskip\\}
{\it
$^{a}$IKP(Theorie), KFA J{\"u}lich, 5170 J{\"u}lich, Germany
\medskip\\
$^{b}$L. D. Landau Institute for Theoretical Physics, GSP-1,
117940, \\
ul. Kosygina 2, Moscow 117334, Russia \medskip\\
$^{c}$Interdisciplinary Laboratory for Advanced Studies (ILAS)\\
Miramare, I-34014 Trieste, Italy
\vspace{1.0cm}\\ }
{\Large
Abstract}\\
\end{center}
We study a novel property of large rapidity-gap events in
deep inelastic scattering at HERA:
splitting the pomeron into two jets in the
photon-pomeron fusion reaction
$\gamma^{*}\Pom \rightarrow q\bar{q}$.
It gives rise to the diffraction dissociation of virtual
photons into the back-to-back jets.
We find that at large invariant mass   $M$ of two jets,
$M^{2}\gg  Q^{2}$, the transverse momentum of jets comes
from the intrinsic transverse momentum of gluons in the pomeron,
and the photon-pomeron fusion
direcly  probes the diffrential gluon structure
function of the proton $\partial G(x,q^{2})/\partial \log q^{2}$ at
the virtuality $q^{2}\sim k^{2}$.
We present estimates for the jet production cross section,
which show the process is easily measurable at HERA.
 \bigskip\\

\begin{center}
E-mail: kph154@zam001.zam.kfa-juelich.de
\end{center}

\pagebreak

Diffraction dissociation reactions $a+p \rightarrow X +p'$ can be
interpreted as interaction of the projectile $b$ with the pomeron
$a+\Pom \rightarrow X$ [1]. When the projectile is the virtual
photon $\gamma^{*}$, then at moderately large
invariant mass $M$ of the diffractively
excited state, $M^{2}\sim Q^{2}$, one probes
the valence $q\bar{q}$ component of the pomeron ($Q^{2}$ is the
virtuality of the photon). The corresponding
valence structure function of the
pomeron was evaluated in [2] and the predicted rate of the
diffraction dissociation is consistent with the first data from
the ZEUS experiment at HERA [3].

In [2] we have also predicted diffraction dissociation of photons
into back-to-back jets, alias the splitting of pomerons into two
jets, the partonic subprocess for which is
the (virtual) photon-pomeron fusion
\beq
\gamma^{*}+\Pom \rightarrow q+\bar{q} \, .
\label{eq:1}
\endeq
This process has a very clean signature: the
(anti)quarks produced with large transverse momentum $k$
with respect to the photon-pomeron collision axis,
materialize as the back-to-back jets in the photon-pomeron
c.m.s.
Understanding to which extent the pomeron can be treated as
a particle is one of the pressing issues in the pomeron physics [2,4],
and the salient feature of this "two-body" reaction
(\ref{eq:1}) is precisely that here the pomeron participates as an
"elementary" particle, which carries 100 per cent of its momentum.
On the other hand, the QCD pomeron describes the colour-singlet
exchange by (the two) gluons and, as such, it is definitely not an
elementary particle. In this paper we update the considerations
[2] and discuss in detail how this gluonic substructure of the pomeron
manifests itself in the back-to-back jet production. Our major
finding is that, in the most interesting kinematical domain,
the reaction (\ref{eq:1}) proceeds via interaction of the photon
with the two gluons of the pomeron in such a way
that the transverse momenta of the quark and antiquark jets
come entirely from the intrinsic
transverse momentum of gluons in the
pomeron. Consequently, the underlying process can be dubbed
the splitting of pomerons into two jets. We demonstrate how
the diffraction dissociation of photons into the back-to-back
jets direcly probes the differential gluon structure function
of the proton $f(x,q^{2})=\partial G(x,q^{2})/\partial \log q^{2}$ at
the virtuality $q^{2}\sim k^{2}$.

Experimentally, one studies the reaction $\gamma^{*}+p\rightarrow
X+p$, which can be viewed as emisssion of the pomeron $\Pom$
by the proton followed by the photoabsorption $\gamma^{*}+\Pom
\rightarrow X$. By definition of the diffraction dissociation,
the fraction $x_{\Pom}$ of proton's momentum carried by the
pomeron is small: $x_{\Pom} \lsim$ (0.05-0.1). The recoil proton
emerges in the final state separated from the debris of the
diffractively excited photon by a large (pseudo)rapidity gap
\beq
\Delta \eta \approx \log\left({1\over x_{\Pom}}\right)\, .
\label{eq:2}
\endeq
The invariant mass M of the diffractively excited state and
the total mass squared $s$ of the hadronic final state are
related by $M^{2}+Q^{2} = x_{\Pom}(s+Q^{2})$.
Hereafter, for the sake of simplicity, we consider the forward
diffraction dissociation, with the vanishing transverse momentum
$\vec{p}_{\perp}$ of the recoil proton: $t=-\vec{p}_{\perp}^{2}
=0$; extension of our results to finite, but small $t\ll k^{2}$
poses no problems. We consider the jet cross section at the partonic
level,
i.e., we calculate the differential cross section for production
of the high-$k$ (anti)quarks in the photon-pomeron fusion
(\ref{eq:1}). The diffraction dissociation cross section is
described by diagrams of Fig.1.

The staring point of our analysis
is the generalization of formulas for the corresponding
differential cross section $d\sigma_{D}(\gamma^{*}_{T,L}\rightarrow
q\bar{q})/dt$ derived by us in [2,5]. For the diffraction dissociation
of the ($\gamma_{T}^{*}$) transverse and ($\gamma_{L}^{*}$)
longitudinal photons we obtain
\arr
\left.{d\sigma_{D}(\gamma_{T}^{*}\rightarrow
q\bar{q})\over dt}\right|_{t=0}=
{\alpha_{em}\over 6\pi}\sum e_{f}^{2} \int_{0}^{1} dz~
\int d^{2}\vec{k}~d^{2}\vec{\kappa}~d^{2}\vec{\kappa}'~
\alpha_{S}^{2}\cdot
{1\over \kappa^{4}}{\partial G(x_{g},\kappa^{2})\over \partial
\log \kappa^{2}} \cdot
{1\over {\kappa'}^{4}}{\partial G(x_{g},{\kappa'}^{2})\over \partial
\log {\kappa'}^{2}} \nonumber \\
\left\{
{[z^{2}+(1-z)^{2}]\vec{k}^{2}+m_{f}^{2} \over
[\vec{k}^{2}+\varepsilon^{2}]^{2} } -
{[z^{2}+(1-z)^{2}]\vec{k}\cdot(\vec{k}+\vec{\kappa})+m_{f}^{2} \over
[\vec{k}^{2}+\varepsilon^{2}]
[(\vec{k}+\vec{\kappa})^{2}+\varepsilon^{2}] }\right.~~~~~~~~~~~~~~~ \\
\left.
-{[z^{2}+(1-z)^{2}]\vec{k}\cdot(\vec{k}-\vec{\kappa}')+m_{f}^{2} \over
[\vec{k}^{2}+\varepsilon^{2}]
[(\vec{k}-\vec{\kappa}')^{2}+\varepsilon^{2}] } +
{[z^{2}+(1-z)^{2}]
(\vec{k}+\vec{\kappa})\cdot(\vec{k}-\vec{\kappa}')+m_{f}^{2} \over
[(\vec{k}+\vec{\kappa})^{2}+\varepsilon^{2}]
[(\vec{k}-\vec{\kappa}')^{2}+\varepsilon^{2}] } \right\} \, ,
\nonumber
\label{eq:3}
\endarr
\arr
\left.{d\sigma_{D}(\gamma_{L}^{*}\rightarrow
q\bar{q})\over dt}\right|_{t=0}=
{\alpha_{em}\over 6\pi}\sum e_{f}^{2} \int_{0}^{1} dz~4Q^{2}
z^{2}(1-z)^{2}\int
d^{2}\vec{k}~d^{2}\vec{\kappa}~d^{2}\vec{\kappa}'~\nonumber \\
\alpha_{S}^{2}\cdot
{1\over \kappa^{4}}{\partial G(x_{g},\kappa^{2})\over \partial
\log \kappa^{2}} \cdot
{1\over {\kappa'}^{4}}{\partial G(x_{g},{\kappa'}^{2})\over \partial
\log {\kappa'}^{2}}
\left\{
{1 \over
[\vec{k}^{2}+\varepsilon^{2}]^{2} } -
{1 \over
[\vec{k}^{2}+\varepsilon^{2}]
[(\vec{k}+\vec{\kappa})^{2}+\varepsilon^{2}] }\right. \nonumber \\
\left.
-{1 \over
[\vec{k}^{2}+\varepsilon^{2}]
[(\vec{k}-\vec{\kappa}')^{2}+\varepsilon^{2}] } +
{1 \over
[(\vec{k}+\vec{\kappa})^{2}+\varepsilon^{2}]
[(\vec{k}-\vec{\kappa}')^{2}+\varepsilon^{2}] } \right\}  \, .
\label{eq:4}
\endarr
Here $e_{f}$ is the quark charge in units  of the electron charge,
$m_{f}$ is the quark mass,
 $z$ is a fraction of the photon's  lightcone momentum
carried by the (anti)quark,
and $\vec{k},\vec{\kappa},\vec{\kappa}'$ are the
transverse momenta
of the produced quark and of gluons in pomerons (Fig.1).
The observed back-to-back jets
will have the transverse momenta $\vec{k}$ and $-\vec{k}$.
The invariant mass $M$ of the
two-jet final state equals
\beq
M^{2}={m_{f}^{2}+\vec{k}^{2} \over z(1-z) }
\label{eq:5}
\endeq
and
\beq
dz\, dk^{2} = dM^{2} \, dk^{2} {m_{f}^{2}+k^{2}\over M^{4}}
\left(1-4{m_{f}+k^{2}\over M^{2}}\right)^{-1/2}\,   ,
\label{eq:6}
\endeq
where the factor
\beq
{1\over \cos\theta}=
\left(1-4{m_{f}+k^{2}\over M^{2}}\right)^{-1/2}
\label{eq:7}
\endeq
corresponds to the familiar Jacobian-peak at the jet production
angles $\theta \sim 90^{o}$ in the photon-pomeron c.m.s.
In the Born approximation, the differential gluon density is
related to the two-body formfactor of the nucleon
$\langle N|\exp(i\vec{\kappa}_{1}\vec{r}_{1}+i\vec{k}_{2}\vec{r}_{2})
|N\rangle$ by the equation [6]
\beq
{1\over \kappa^{4}}
{\partial G(x,\kappa^{2})\over \partial \log \kappa^{2}}=
{4\alpha_{S}(\kappa^{2})\over \pi (\kappa^{2}+\mu_{G}^{2})^{2}}
\left(1-
\langle N|\exp(i\vec{\kappa}(\vec{r}_{1}-\vec{r}_{2})
|N\rangle \right)\, ,
\label{eq:8}
\endeq
and the limiting form of Eqs.~(3,4)
derived in [2,4] is obtained. (In (\ref{eq:8}), $\mu_{G}=R_{c}^{-1}$
and $R_{c}$ is the correlation radius for the perturbative gluons
[2,4-9]).

After integrations over the azymuthal angles of the gluon momenta
$\vec{\kappa}$ and $\vec{\kappa}'$, the differential cross sections
can be written in the compact form
\arr
\left.{d\sigma_{D}(\gamma_{T}^{*}\rightarrow q\bar{q}) \over
dM^{2}dk^{2}dt}\right|_{t=0}={\pi^{2} \alpha_{em} \over 6 \cos\theta}
\cdot {m_{f}^{2}+k^{2} \over M^{4}}\alpha_{S}^{2}(k^{2})
\left\{ \left(1-2{k^{2}+m_{f}^{2}\over M^{2}}\right)
\Phi_{1}^{2} + m_{f}^{2}\Phi_{2}^{2}
\right\}  \, ,
\label{eq:9}
\endarr
\arr
\left.{d\sigma_{D}(\gamma_{L}^{*}\rightarrow q\bar{q}) \over
dM^{2}dk^{2}dt}\right|_{t=0}={\pi^{2} \alpha_{em} Q^{2}
\over 6 \cos\theta}
\cdot {(m_{f}^{2}+k^{2})^{3} \over M^{8}}\alpha_{S}^{2}(k^{2})
\Phi_{2}^{2}  \, ,
\label{eq:10}
\endarr
where
\arr
\Phi_{1}=\int {d\kappa^{2}\over \kappa^{4}}
f(x,\kappa^{2})
\left[{k\over k^{2}+\varepsilon^{2}}-
{k\over \sqrt{a^{2}-b^{2}}}+
{2k\kappa^{2} \over a^{2}-b^{2}+a\sqrt{a^{2}-b^{2}}}\right]
 \nonumber \\
= {2k\varepsilon^{2}\over (k^{2}+\varepsilon^{2})^{3} }
\int {d\kappa^{2}\over \kappa^{2}}
W_{1}({k^{2}\over \varepsilon^{2}},{\kappa^{2}\over \kappa^{2}})
f(x,\kappa^{2})\, ,~~~~~~~~~~~~~
\label{eq:11}
\endarr

\beq
\Phi_{2}=\int {d\kappa^{2}\over \kappa^{4}}
f(x,\kappa^{2})
\left[
{1\over \sqrt{a^{2}-b^{2}}} -
{1\over k^{2}+\varepsilon^{2}}
\right] \approx
 {k^{2}-\epsilon^{2}\over (k^{2}+\varepsilon^{2})^{3} }\cdot
\int {d\kappa^{2}\over \kappa^{2}}
W_{2}({k^{2}\over \varepsilon^{2}},{\kappa^{2}\over \kappa^{2}})
f(x,\kappa^{2})\, ,
\label{eq:12}
\endeq
\beq
a=\varepsilon^{2}+k^{2}+\kappa^{2} \, ,
\label{eq:13}
\endeq
\beq
b=2k\kappa\, .
\label{eq:14}
\endeq
The useful relations are also
\beq
\varepsilon^{2}=(k^{2}+m_{f}^{2}){Q^{2}\over M^{2}}+m_{f}^{2} \, ,
\label{eq:15}
\endeq
\beq
k^{2}+\varepsilon^{2}=(k^{2}+m_{f}^{2}){M^{2}+Q^{2}\over M^{2}} \, .
\label{eq:16}
\endeq

The kernels $W_{i}$ are functions of the dimensionless variables
$\omega = k^{2}/\varepsilon^{2}$ and $\tau =\kappa^{2}/k^{2}$.
They have similar properties as a function of $\omega$ and $\tau$.
Hereafter we concentrate on diffraction dissociation of
the transverse photons, because diffraction dissociation
of the longitudinal photons into two jets has much smaller
cross section [2]. We also can neglect the $m_{f}^{2}\Phi_{2}^{2}$
term in Eq.~(\ref{eq:9}).

The kernel $W_{1}$ is shown in Fig.~2. In deep inelastic
scattering, at large $Q^{2}$, we have $\omega = M^{2}/Q^{2}$.
In the excitation of
the moderately large masses $M^{2}\lsim Q^{2}$, the kernel
$W_{1}$ is essentially flat, has the unity height, and extends up
to $\kappa^{2}\approx A_{T}k^{2}$, where $A_{T}\approx 3 $.
Consequently, at $\omega \lsim 1$,
\beq
\int d\tau \,
W_{1}(\omega,\tau)
f(x,\kappa^{2}) = G(x,A_{T}k^{2})
\label{eq:17}
\endeq
and
\beq
\Phi_{1}={2k\varepsilon^{2}\over (k^{2}+\varepsilon^{2})^{3} }
G(x,A_{T}k^{2})\, ,
\label{eq:18}
\endeq
which updates Eq.~(60) of [2].
The corresponding differential cross section for the jet
production at $M^{2}\lsim Q^{2}$ equals
(here we neglect the last $m_{f}^{2}$ term in $\varepsilon^{2}$
in the numerator)
\arr
\left.{d\sigma_{D}(\gamma_{T}^{*}\rightarrow q\bar{q}) \over
dM^{2}dk^{2}dt}\right|_{t=0}=
\sum e_{f}^{2}
{\pi^{2} \alpha_{em}
\alpha_{S}^{2}(k^{2})
\over 6 \cos\theta}
\left(1-2{k^{2}+m_{f}^{2}\over M^{2}}\right)  \\
\cdot {1 \over M^{4}}
{4k^{2}\over (k^{2}+m_{f}^{2})^{3} }
\left({Q^{2}\over M^{2}+Q^{2}}\right)^{2}
\left({M^{2}\over M^{2}+Q^{2}}\right)^{3}
G(x,A_{T}k^{2})^{2}\nonumber
\label{eq:19}
\endarr
Because of the logarithmic $\kappa^{2}$ integration, at
$\omega \lsim 1$ the major
contribution to the integral for $\Phi_{1}$ comes from
$\kappa^{2}\ll k^{2}$ and we have a semblance of the
Leading-Log$Q^{2}$ approximation (LLQA).
 This means that the transverse momentum
of gluons of the pomeron contributes little to the transverse
momentum of jets. Therefore, in the diffractive excitation to
low mass states $M^{2}\lsim Q^{2}$,
the transverse momentum of jets comes predominantly from the
intrinsic transverse momentum of the quark and antiquark in the
$q\bar{q}$ Fock state of the photon.
The $dk^{2}/k^{4}$ transverse momentum distribution corresponds
to a dominance of the diffraction dissociation cross section
by the $q\bar{q}$ configurations in which (anti)quarks have
small, hadronic, value of the transverse momentum irrespective
of the value of $Q^{2}$ [2,4].

Much more interesting case is diffraction excitation of heavier masses,
$M^{2} \gg Q^{2}$, i.e., $\omega \gg 1$,  which requires
utilization of the transverse momentum  of gluons in the pomeron.
Here the kernel develops a very sharp resonance peak
at $\tau=\kappa^{2}/ k^{2} \sim 1$, which
clearly shows that in
this regime the transverse momentum of jets comes
entirely from the transverse momentum $\vec{\kappa} \approx
\vec{k}$ of gluons. The LLQA considerations are completely
inapplicable here.
This resonance contribution can be quantified as follows:
The height of the peak approximately follows the law
\beq
W_{1}^{(max)} \sim {\omega \over 2}\, ,
\label{eq:20}
\endeq
and the width of the peak is approximately constant in the
$\log(\kappa^{2}/k^{2})$ scale. A separation of the kernel
$W_{1}$ into the peak component $P_{1}$ and the plateau is
somewhat convention dependent. We define the resonance peak
distribution via the decomposition
\beq
W_{1}(\omega,\tau) = W_{1}(\omega=1,\tau)+ {\omega\over 2}
P_{1}(\omega,\tau) \,.
\label{eq:21}
\endeq
The resulting peak distribution function $P_{1}(\omega,\tau)$
and its variation with $\omega$ are shown in Fig.~3. At
$\omega >> 1$ it becomes an approximately
scaling function of $\tau$. Because $P_{1}(\omega,\tau)$
is much sharper function of $\kappa^{2}$ than $f(x,\kappa^{2})$,
we can write
\beq
\int  d\log \tau \,
P_{1}(\omega,\tau)
f(x,\kappa^{2}) = S(\omega)f(x,\tau_{T}k^{2})\, ,
\label{eq:22}
\endeq
where $S(\omega)$ is an area under the resonance peak, and
$\tau_{T}$ corresponds to the center of gravity of the peak,
$\langle \log \tau \rangle =\log \tau_{T}$. The
area under the peak $S(\omega)$ and its center of gravity
are shown in Fig.~4.
At the asymptotically large $\omega \gg 1$ the resonance
area integrates to unity, $S(\omega \gg 1) =1$, which is
a viable first approximation at $\omega \gsim 3$.
Consequently, at $\omega \gg 1$
\arr
\Phi_{1}=
{k\over (k^{2}+\varepsilon^{2})^{2} }\left[
{2G(x,A_{T}k^{2})\over \omega }+
S(\omega)f(x,\tau_{T}k^{2})\right] \approx
{k\over (k^{2}+\varepsilon^{2})^{2} }
S(\omega)f(x,\tau_{T}k^{2})
\label{eq:23}
\endarr
and
\arr
\left.{d\sigma_{D}(\gamma_{T}^{*}\rightarrow q\bar{q}) \over
dM^{2}dk^{2}dt}\right|_{t=0}=~~~~~~~~~~~~~~~~~~~~~~~~~~~~~~~\\
\sum e_{f}^{2}
{\pi^{2} \alpha_{em}
\alpha_{S}^{2}(k^{2})
\over 6 \cos\theta}
\left(1-2{k^{2}+m_{f}^{2}\over M^{2}}\right)
\cdot {1\over M^{4}}
{k^{2}\over (k^{2}+m_{f}^{2})^{3}}
\left({M^{2}\over M^{2}+Q^{2}}\right)^{3}
f(x,\tau_{T}k^{2})^{2} \, ,\nonumber
\label{eq:24}
\endarr
which update Eqs.~(57,58) of [2].
The emerging possibility of directly measuring the  differential gluon
distribution is a novel property of the pomeron splitting reaction.
For the numerical estimates of $f(x,Q^{2})$ in terms of the more
familiar integrated gluon distribution function $G(x,Q^{2})$,
we can use the so-called Double-Leading-Log-Approximation (DLLA)
identity, which also holds in the BFKL regime [7] (here
$N_{f}=3$ active flavours are assumed)
\beq
{3\over 4}\cdot {1\over G(\xi,Q^{2})}
{\partial^{2} G(\xi,r)\over \partial \log(1/x)\,\partial
 \log \log Q^{2}} =1\, .
\label{eq:25}
\endeq
In [7,8] we have shown that the effective intercept
\beq
\Delta_{eff}(x,Q^{2})= {1\over G(x,Q^{2})}
{\partial G(x,Q^{2}) \over \partial \log(1/x)}
\label{eq:26}
\endeq
is a relatively slow function of $Q^{2}$ and $x$, and in the
kinematical range of the interest at HERA, $\Delta_{eff}(x,Q^{2})
\approx \Delta_{\Pom}=0.4$ to better than
the factor 2. Here $\Delta_{\Pom}$
is an intercept of the BFKL pomeron evaluated in [8]. Then the DLLA
identity (\ref{eq:25}) gives an estimate
\beq
f(x,Q^{2}) \approx {3\alpha_{S}(k^{2}) \over \pi\Delta_{\Pom}}
G(x,Q^{2})\,.
\label{eq:27}
\endeq
This estimate shows that the term $\propto f(x,\xi_{T}k^{2})$ in
Eq.~(\ref{eq:23}) dominates already starting with $\omega \gsim
(2-4)$. Only this term contributes to the real photoproduction,
$Q^{2}=0$, at $k^{2}\gg m_{f}^{2}$ of the interest.

Above we have considered the forward diffraction dissociation $t=0$.
At $|t|\ll k^{2}$, which is the dominant region in
diffraction dissociation into two jets, we can write
$d\sigma_{D}/dt =d\sigma_{D}/dt|_{t=0}\cdot \exp(B_{D}t)$ with
the slope $B_{D}$ close to the slope of the $\pi N$ elastic
scattering, $B_{D}\sim 10$GeV$^{-2}$. The cross section, integrated
over the recoil protons, can be estimated as
\beq
\int dt {d\sigma_{D}(\gamma^{*}\rightarrow q+\bar{q})
\over dM^{2}dk^{2}dt}\approx
{1\over B_{D}}\left.
{d\sigma_{D}(\gamma^{*}\rightarrow q+\bar{q})
\over dM^{2}dk^{2}dt}\right|_{t=0}
\label{eq:28}
\endeq
In order to have an  idea on the magnitude of the pomeron splitting
cross section in the real photoproduction, consider the cross
section  integrated over masses $M^{2}\gsim 4k^{2}$. Ignoring
certain enhancement  coming from the Jacobian peak, and
putting $\cos\theta \sim 1$, we find
\beq
{d\sigma_{D} \over dk^{2}} \approx
{\pi^{2} \alpha_{em}
\alpha_{S}^{2}(k^{2})
\over 24 B_{D}}
{1\over (k^{2}+m_{f}^{2})^{3}}
f(x,\tau_{T}k^{2})^{2} \, ,
\label{eq:29}
\endeq
which is dominated by $M^{2} \sim 4k^{2}$.
The total cross section for producing jets with the transverse
momentum above certain threshold $k^{2}> k_{0}^{2}$ equals
\beq
\sigma_{D}(k^{2}>k_{0}^{2}) =\int_{k_{0}^{2}} dk^{2}
{d\sigma_{D} \over dk^{2}} \approx
{\pi^{2} \alpha_{em}
\alpha_{S}^{2}(k_{0}^{2})
\over 48 B_{D}}
{1\over (k_{0}^{2}+m_{f}^{2})^{2}}
f(x,\tau_{T}k_{0}^{2})^{2} \, .
\label{eq:30}
\endeq
Take $k_{0}^{2}=5$GeV$^{2}$ in the real photoproduction at the
typical HERA center of mass energy squared
$s=4\nu E_{p} \sim 5\cdot 10^{4}$GeV$^{2}$. The gluon distribution
must be evaluated at $x=x_{0}\sim M^{2}/s \sim 4k_{0}^{2}/s
\approx 5\cdot 10^{-4}$. Our evaluation [9] based on the
BFKL evolution analysis, gives $G(x_{0}, \tau_{T}k_{0}^{2})
\sim 20$ and Eq.~(\ref{eq:27}) gives $f(x_{0},\tau_{T}k_{0}^{2})
\sim 20$. Then $\sigma_{D}(k^{2}\gsim 5{\rm GeV}^{2}) \sim
0.1 \mu$b. This must be compared with
our prediction [2] that the
diffraction dissociation of real photons into the $q\bar{q}$
states makes about 15\% of the total cross section, i.e.,
$\sigma_{D}(tot)\sim 20\mu$b, which is in perfect agreement with the
preliminary data from HERA [10].

We conclude that diffraction dissociation of photons into
back-to-back jets has sufficiently large cross section to be
observable at HERA, and is a potentially
important tool for measuring the differential gluon distribution
$\partial G(x,Q^{2})/\partial \log Q^{2}$.

A fresh look at the difference of mechanisms of production of
large transverse momentum jets at $M^{2} \lsim Q^{2}$ and
$M^{2}\gg Q^{2}$ is in order. Usually, one thinks of
production of jets in terms of the elementary partonic subprocesses
like the photon-gluon fusion $\gamma^{*} + g \rightarrow q+\bar{q}$
with production of the large-$k$ quarks. We have shown that in
the regime of $M^{2}\lsim Q^{2}$, gluons in the pomeron have small
transverse momenta $\kappa^{2}\ll k^{2}$ and, with certain
reservations, here pomeron acts like an elementary particle
whose substructure is not important. In the opposite to that,
at $M^{2}\gg Q^{2}$, the transverse momenta of the back-to-back
jets rather come from the back-to-back transverse momentum of
gluons in the pomeron, and the relevant subprocess for splitting
the pomeron into two jets is the manifestly three-body,
photon-two-gluon collision in the initial state.
Here the gluonic substructure
of the pomeron enters in a most crucial manner, which precludes
any consideration of the pomeron as an elementary particle.
Furthermore, it would be completely illegitimate to interpret
such a jet production as interaction of the photon whith
one parton of the pomeron which carries 100\% of the pomeron's
momentum. Notice, that in the regime of $M^{2}\lsim Q^{2}$
the cross section
is $\propto G(x,k^{2})^{2}$, whereas at $M^{2}\gg Q^{2}$
it is
$\propto [\partial G(x,k^{2})/\partial \log k^{2}]^{2}$. This
striking difference between the two cases
is just one of manifestations of the lack of factorization
for the QCD pomeron contribution to the diffraction
dissociation [2].
\medskip\\
{\bf \Large Conclusions:\smallskip\\}
We found a novel mechanism of diffraction dissociation of
photons into two back-to-back jets, in which the transverse
momentum of jets
originates from the transverse momentum of gluons in the pomeron.
This mechanism can not be interpreted in familar terms of
hard interaction of the photon with one energetic parton of
the pomeron.
The corresponding jet production cross section Eq.~(\ref{eq:24})
emerges as a unique direct probe of the differential gluon
structure function of the proton $\partial G(x,Q^{2})
/\partial \log Q^{2}$. The jet production cross section is
large and must easily be measurable at HERA. \medskip\\

{\bf Acknowledgements}: B.G.Z. is grateful to
J.Speth for the hospitality at IKP, KFA J\"ulich, where this
work was initiated and to S.Fantoni for the hospitality at
SISSA, where this work was finished.
\pagebreak

{\bf Figure captions:}

\begin{itemize}
\item[Fig.1 - ]
One of the 16
perturbative QCD diagrams for the diffraction dissociation
cross section. The vertical dashed line shows the unitarity
cut corresponding to the diffractively produced state.

\item[Fig.2 - ]
The kernel $W_{1}(\omega,\tau)$ for different values of
$\omega$. The resonance peak at $\tau \sim 0$ developes
at large $\omega$.

\item[fig.3 - ]
The scaling properties of the resonanace peak $P_{1}(\omega,\tau)$.

\item[Fig.4 - ]
The area of the resonance peak $S(\omega)$ and its
center of gravity $\tau_{T}$ as a function of $\omega$.

\end{itemize}
\end{document}